\documentstyle[epsfig,mathptm,twocolumn]{mn} 
\newif\ifAMStwofonts
\AMStwofontstrue                
\DeclareMathVersion{bold}                    
\DeclareMathAlphabet{\mathbfit}{OT1}{cmr}{bx}{it}
\SetMathAlphabet\mathbfit{bold}{OT1}{cmr}{bx}{it}
\DeclareMathAlphabet{\mathbfss}{OT1}{cmss}{bx}{n}
\SetMathAlphabet\mathbfss{bold}{OT1}{cmss}{bx}{n}
\ifAMStwofonts
  \ifCUPmtlplainloaded \else
    \DeclareSymbolFont{UPM}{U}{eur}{m}{n}
    \SetSymbolFont{UPM}{bold}{U}{eur}{b}{n}
    \DeclareSymbolFont{AMSa}{U}{msa}{m}{n}
    \DeclareMathSymbol{\upi}{0}{UPM}{"19}
    \DeclareMathSymbol{\umu}{0}{UPM}{"16}
    \DeclareMathSymbol{\upartial}{0}{UPM}{"40}
    \DeclareMathSymbol{\leqslant}{3}{AMSa}{"36}
    \DeclareMathSymbol{\geqslant}{3}{AMSa}{"3E}

     \let\le=\leqslant

  \fi
\fi

\newcommand{\mnras}{{MNRAS}}

\newcommand{\apj}{{ ApJ}}
\newcommand{\apjl}{{ ApJL}}
\def\kms{\ifmmode {\rm \ km \ s^{-1}}\else$\rm km s^{-1}$\fi}
\def\dxo{{\rm d} {\bf x}^{\rm o}}

\def\P{{\rm Pr}}
\def\A{{\mathcal A}}
\def\xo{{\bf x}^{\rm o}}
\def\xp{{\bf x}^{\rm p}}
\def\xf{{\bf x}^{\rm f}}

\def\xb{{\bf x}^b}
\def\xpp{{{\bf x}^{\rm p}}^{\prime}}
\def\d{\rm d}
\def\N{{\mathcal N}}
\def\M{{\mathcal M}}

\def\Fb{{\bf {F}}_b} 
\def\Fbb{{\bf {F}}_{\bar{b}}} 
\def\sigc{\sigma_{\rm c}}
\def\sigb{\sigma_{\rm b}}
\def\gs{\mathrel{\raise0.35ex\hbox{$\scriptstyle >$}\kern-0.6em 
\lower0.40ex\hbox{{$\scriptstyle \sim$}}}}
\def\ls{\mathrel{\raise0.35ex\hbox{$\scriptstyle <$}\kern-0.6em 
\lower0.40ex\hbox{{$\scriptstyle \sim$}}}}

\title
[Analytic marginalization over CMB calibration and beam uncertainty]
{Analytic marginalization over CMB calibration and beam uncertainty}
\author[S.L.~Bridle et al.]
{S.L.~Bridle$^1$, R. Crittenden$^2$, A.~Melchiorri$^3$, M.P.~Hobson$^4$, R.~Kneissl$^4$, A.N.~Lasenby$^4$\\
$^1$Institute of Astronomy, University of Cambridge, Madingley Road,
    Cambridge CB3 0HA, UK\\
$^2$ DAMTP, Wilberforce Road, Cambridge CB3 0HA, UK\\
$^3$ Nuclear and Astrophysics Laboratory, 1 Keble Road, Oxford OX1 3RH, UK\\
$^4$ Astrophysics Group, Cavendish Laboratory,  Madingley Road, 
Cambridge CB3 0HE, UK\\
}

\date{\today}
\pagerange{\pageref{firstpage}--\pageref{lastpage}}
\pubyear{2001}
\begin{document}
\maketitle
\label{firstpage}
\begin{abstract} 
With the increased accuracy and angular scale coverage of the recent 
CMB experiments it has become important to include
calibration and beam uncertainties when estimating cosmological parameters.
This requires an integration over possible values of the calibration and
beam size, which can be performed numerically but greatly increases 
computation times.
We present a fast and general method for marginalization 
over calibration-type errors by analytical integration.
This is worked through for the specific example of CMB
calibration and beam uncertainties and the resulting formulae
are practical to implement.
We show how cosmological parameter constraints from the latest CMB
data are changed when 
calibration/beam uncertainties are taken into account:
typically the best fit parameters are shifted and the 
errors bars are increased by up to fifty per cent
for e.g. $n_s$ and $\Omega_{\rm b} h^2$; although as expected 
there is no change for $\Omega_{\rm K}$, because it is constrained by 
the \emph{positions} of the peaks.
\end{abstract}


\section{Introduction}
\label{intro}

Fluctuations in the cosmic microwave background (CMB) radiation on scales of
fractions of a degree and larger are potentially a direct probe of the state
of the universe 300,000 years after the big bang, modified by the geometry
of the universe. If the initial fluctuations were Gaussian and structure 
formed by gravitational collapse then the angular power spectrum of the CMB
contains much cosmological information, and is also easy to calculate 
using codes CMBFAST (Seljak \& Zaldarriaga 1999) and CAMB (Lewis, Challinor 
\& Lasenby 1999). Therefore many experiments have been carried out to
estimate its form, and the results from the second generation of CMB 
telescopes are eclipsing previous results. Recent work involving 
parameter estimation from the CMB includes 
Wang, Tegmark \& Zaldarriaga (2001),
Netterfield el (2001), de Bernardis et al. (2001), 
Pryke et al. (2001), Stompor et al. (2001),
Jaffe et al. (2001),
Bridle et al. (2001), 
Kinney, Melchiorri \& Riotto (2001),
Le Dour et al. (2000), 
Lahav et al. (2000), 
Dodelson \& Knox (2000),
Melchiorri et al. (2000),
Efstathiou (2000),
Gawiser \& Silk (1998)
and Lineweaver (1998).

CMB power spectrum results have significant calibration uncertainties, 
due either to 
uncertainty in the 
flux of the calibration
source (e.g. Jupiter)
or difficulties in its measurement
by the experiment in question (e.g. 
a scan synchronous noise in measurements of the CMB dipole) or both.
As a result the band power $\Delta T$ or $\Delta T^2$ estimates 
from any single experiment can be scaled up or down by some unknown 
factor. This calibration uncertainty is now of greater significance 
because of the increased precision of experiments: it is now of a 
similar size to the quoted random errors. In addition, because of the 
correlations in errors that it introduces, the calibration 
uncertainty is not simple to take into consideration
when extracting cosmological parameters from a CMB power spectrum.
This contrasts 
with the case of a calibration uncertainty on a single data point, 
which can be approximately taken into account by adding the 
calibration uncertainty in quadrature with the random errors.

A fast method in the literature for dealing with this uncertainty 
couples the marginalization over the calibration with that 
over the CMB power spectrum normalization (Ganga et al. 1997, 
Lange et al. 2001). However, this is non-trivial to extend to the 
case where several CMB data sets have independent and significant
calibration uncertainties ($\sim 20$ per cent in ${\rm d}T^2$ 
for  BOOMERANG, Netterfield et al. 2001, and $\sim 8$ per cent
for MAXIMA-1, Lee et al. 2001, and DASI, Halverson et al 2001). 
Wang et al. (2001) account for the calibration uncertainty by 
using a method related to that presented here, however the derivation 
is not well documented in the literature.
Frequently the marginalisation is carried out numerically, which
is time consuming.
Here we present the full derivation of a fast
method, in which the calibration correction for a single data set 
is marginalised over analytically.
This takes no 
more computation time than 
when the
calibration uncertainty is ignored.

In addition to a calibration uncertainty, the experiments
such as BOOMERANG and MAXIMA suffer from beam uncertainties.
Many systematics can produce an uncertainty in the reconstruction of
the instrumental beam. For example, in the case of the BOOMERANG 
experiment the main contribution to the beam uncertainty is due to the
pointing uncertainty.
Pointing uncertainty leads to beam type uncertainties because 
temperature measurements are not made continually, but instead are
averaged over some discrete time period. High frequency jitter in the
pointing means that each discrete measurement includes pointings over a
region of the sky.  
The effective beam of an experiment reflects the area of the sky being
sampled in a given time step, so the true beam must be convolved with the
area sampled in a given time step.  Thus, jitter translates into a larger
effective beam and uncertainty in the amount of jitter can be treated as
uncertainty in the effective beam size.
The impact on the CMB power spectrum is to introduce an 
`angular-scale-dependent' error.
For conciseness in the rest of this paper, we refer to the combined
effects of pointing and beam uncertainties simply as beam uncertainty.

To include both calibration and beam uncertainties is computationally
costly. 
In general the beam uncertainty is integrated over numerically, thus
the computation time is increased by a factor equal to the number
of integration steps used.
In this paper we show how an `angular-scale-dependent'
uncertainty such as the beam error
can be marginalised over analytically assuming a Gaussian 
prior on the size of the correction. The combined analytic 
calibration and beam marginalisation takes two to three times as
long to calculate as when there are no such uncertainties, and
has already been used in Lahav et al. (2001) and Bean et al. (2001).

This method follows the general approach of marginalizing over
nuisance parameters analytically discussed, for example, in
Gull (1989), Sivia (1996) and Lahav et al. (2000). 
As discussed in this introduction, this work is motivated
by the recent CMB data sets. 
However, the fast marginalisation over a calibration type parameter
should be much more widely applicable in astronomy.
Therefore in Section \ref{marg} we show the analytic marginalised result 
for the general case of a correlated beam uncertainty, 
reserving the mathematical derivation for the Appendix. 
Section \ref{practical} gives fast computational versions of the
formulae for the special cases of CMB calibration alone and CMB
calibration and beam uncertainties.
In Section \ref{app} we illustrate the effect of the marginalisation by
applying them to the latest CMB data.

\section{Analytic Marginalization}
\label{marg}

\subsection{General approach} 

Before addressing the specific problems of marginalizing over calibration
and beam uncertainties, let us examine the more general problem of 
`nuisance' parameters.  Quite often, observational results will depend
on parameters of the measurement which are not precisely known and whose 
value is not of great interest to us.  In such cases, it is useful to determine 
a likelihood function which folds in the effects of the uncertainties in 
these nuisance parameters by marginalizing over their possible values.  
The resulting likelihood no longer directly depends on the nuisance parameter, 
but its uncertainties become incorporated in the modified data and 
its correlation matrix. 

In many problems, we can model the observational data,  $\xo$, 
by the predictions from an underlying
theory, $\xp$, with a correction arising from some uninteresting
(for the present purposes) nuisance parameter $b$.
Thus, the predictions are modified in some way which depends on how the 
nuisance parameter affects the observations.  
As described below, we can model these corrections
by a factor linear in the parameter times some   
`template' $\xb$, so the predictions for the full observations are  
\begin{eqnarray}
{\xp}^\prime &=& \xp+b\xb. 
\end{eqnarray}
Here, the vector $\xb$ is assumed to be an arbitrary 
function of the predictions, 
but not to depend on the data or its noise correlation matrix. 
The parameter 
$b$ can be thought of as a measure of how far the nuisance parameter 
deviates from its expected mean value. 

As an example consider the calibration uncertainty on the CMB
temperature angular power spectrum. In this case the $\xo$ and
$\xp$ are the observed and predicted $\Delta T^2$ bandpower 
measurements. The predictions could be scaled up and down by 
a factor around 1 therefore $\xb=\xp$.

If the observations are subject to some noise 
drawn from a multivariate Gaussian distribution  
with correlation matrix $\N$, then 
the probability of the data given a model 
(the model's likelihood) is given by
\begin{eqnarray}
\P( \xo | N,\xp,\A,b)
\hspace{-0.2cm}
& = &
\hspace{-0.2cm}
N_{\N} \exp
\left[
-\frac{1}{2} 
\left( \xo - \xpp \right)^{\rm T} \N^{-1} 
\left( \xo - \xpp \right)
\right] \,.
\label{pxgivenb}
\end{eqnarray}
where
$N_{\N}=(2\pi)^{-n/2} |\N|^{-1/2}$ 
and $n$ is the number of data points.
(See the start of Section 3 for a discussion about the validity of
the Gaussian assumption for CMB band power measurements.)
To obtain the likelihood of the data independent
of the calibration type errors, we must marginalise over $b$
\begin{eqnarray}
\P(\xo |\N,\xp,\xb,\,\sigb) 
&=& \int \P(\xo, \, b \,|\N,\, \xp,\, \xb,\,\sigb) \,\d b\\
&=& \int \P(\xo | \N,\, \xp,\, \xb,\, b) \, \P(b|\sigb) \, \d b,
\label{intdb}
\end{eqnarray}
using Bayes theorem and assuming $P(b\,|\N,\, \xp,\,\xb,\,\sigb)=
P(b|\sigb)$.

It is not always clear what form the prior on the calibration parameter $b$ 
should take.
However, if this prior has a simple 
form, then this marginalisation can often be performed analytically. 
Here, we assume the prior is a Gaussian of width $\sigma_b$, 
\begin{equation}
P(b|\sigb)=
\frac{1}{(2\pi)^{1/2}\,\sigb} \exp \left[ -b^2/2\sigb^2 \right].
\label{priorb}
\end{equation}
If we were to instead assume a top hat prior on $b$ then the
analytic marginalisation can still be carried out, but leads to
error functions, and is thus more complicated to implement.

Substituting Eq.~\ref{pxgivenb} and Eq.~\ref{priorb} into 
Eq.~\ref{intdb}, gathering up terms in powers of $b$, completing 
the square and integrating over $b$ (see Appendix), one finds 
\begin{eqnarray}
\P(\xo | \N,\, \xp,\,\xb,\,\sigb)
\hspace{-0.25cm}
&=&
\hspace{-0.25cm}
N_{\M}\,
\,
\exp
\left[
-\frac{1}{2} 
(\xo -\xp)^{\rm T} \M^{-1} (\xo -\xp)
\right]
\label{result}
\\
\M^{-1} &=& \N^{-1} - \frac{\N^{-1}\xb{\xb}^{\rm T}\N^{-1}}
{{\xb}^{\rm T} \N^{-1} \xb + \sigb^{-2}}.
\label{matrixM}
\end{eqnarray}
It is straightforward to show that 
\begin{equation}
\M=\N+ \sigb^2 \xb {\xb}^{\rm T},
\end{equation}
which is the Sherman-Morrison result (see e.g. Press et al. 1992).
Note that Eq.~\ref{result}, \ref{matrixM} 
is an \emph{exact} result, and does not,
for example, rely on a Taylor expansion.
This calculation looks almost the same as when the nuisance parameters are 
not taken into account ($\xpp=\xp$ Eq. \ref{pxgivenb}),
except that the matrix $\N$ has been replaced by the matrix $\M$,
which could be a function of the predicted data 
$\xp$.
Note that, even in the case where correlated errors between data points 
are assumed to be negligible, i.e. $\N^{-1}$ is diagonal, the new matrix 
$\M^{-1}$ is not diagonal. The marginalization effectively introduces 
a correlated error between the data points, adding error to the 
collective modes which are affected by the nuisance parameter. 

This technique has a broad array of applications in the various
stages of CMB analysis, from map making to foreground removal
(see e.g. Bond \& Crittenden, 2001).  
For example, it is used to 
remove the effects of foreground contaminants from maps such as the galaxy.  
In previous applications, however, the templates removed were independent of the
theory, so that the correction to the noise matrix must be done only once.
However, in problems such as calibration marginalization, the templates 
depend on the theoretical predictions, so the effective weight matrix must 
be re-evaluated for each theory.  As we show below, however, this does not
significantly slow down the evaluation of the likelihoods.  
   
Nothing in the above is specific to the CMB, so in fact it
could also be used a variety of applications, such as  
the uncertain normalisation
of the matter power spectrum resulting from Lyman$-\alpha$ forest
measurements (e.g. Croft et al. 2001).

\subsection{Marginalization over multiple parameters} 

Often there are several nuisance parameters which must be 
marginalised over simultaneously.  
The above process can be repeated several times, building up more
complicated matrices $\M$. 
As an example, the CMB bandpower measurements are often subject to 
both calibration and beam uncertainties which must be marginalised over. 
(The result for CMB calibration and beam
uncertainties is given in the following Section.)
This may also be occur where there are several experiments each
with internal and correlated external calibration uncertainties, as
discussed in Knox \& Page (2000), and Tegmark \& Zaldarriaga (2001).
One can expand this set even further, including some simple cosmological 
parameters. In particular, the 
overall theory normalization can be performed in the same way. 

When applying this technique to the marginalization over many 
non-orthogonal parameters simultaneously, it can be useful to use the 
block form of the above result. 
The Woodbury formula (also Press et al. 1992) can be
used
\begin{eqnarray}
(\N + X  X^T)^{-1} 
\hspace{-0.2cm}
&=& 
\hspace{-0.2cm}
\N^{-1} - [\N^{-1} X  
(1 + X^T \N^{-1} X)^{-1} X^T \N^{-1}], 
\end{eqnarray}
where X is a $m\times n$ matrix containing $m$ templates of $n$ elements each.  
Note that this requires only the inversion of a $m \times m$ matrix, where 
$m$ is the number of parameters being marginalised over. 
This formula can be much easier to implement than repeated application
of the Sherman-Morrison result once the number of parameters being
integrated over exceeds two.

In the case where more than one variable is
marginalised over,  the
corresponding normalization factor generalizes to, 
\begin{equation}
N_\M = N_\N \left[ \det(1 + X^T \N^{-1} X )\right]^{-1/2}. 
\end{equation}
Again, this only requires the evaluation of the determinant of an 
$m \times m$ matrix, which is computationally inexpensive 
especially given that 
this matrix must already be inverted.  While sometimes this normalization 
factor can be ignored as an overall prefactor, it cannot be dropped when the 
the templates are theory dependent. 

These techniques have been used in the analysis of the COBE data, 
where there was the possibility that the data were contaminated by a
quadrupole of unknown amplitude. This was accounted for by 
marginalisation over the possible amplitude of the quadrupoles,
by modifying the noise 
correlation matrix using the Woodbury formula (Bond, Jaffe
\& Knox 1998).

\section{Practical Implementation}
\label{practical}

In this Section,  the above equations are rewritten in
a more computationally practical way for the specific case of CMB 
calibration and beam uncertainties in power spectrum measurements.

The analytic technique described above assumes that the observed data
are Gaussianly distributed.  This is clearly only an
approximation for band power measurements, as the 
power spectrum must be positive.
When there are only a few modes in a band power measurement, a
better approximation is to assume a log-normal probability distribution
(Bond, Jaffe \& Knox, 2000).  However, if there are sufficiently many
modes and the error bars are significantly smaller than the measured band
powers, then the Gaussian approximation should be reasonable and the techniques
we describe can be applied. 

\subsection{Calibration Uncertainty}

First we consider the case of a calibration uncertainty in 
the CMB temperature anisotropy angular
power spectrum. The data $\xo$ are the CMB $\Delta T^2$ bandpower
measurements and the $\xp$ are the predicted bandpowers.
In the case of a calibration uncertainty, the predictions
are those expected from the underlying model, multiplied by
some unknown factor, $c$ which is the potentially incorrect
experimental calibration, thus the predictions become
$\xpp = c\xp$.
For example, BOOMERANG has a calibration uncertainty
of $20$ per cent in $\Delta T^2$ 
and
therefore $c$ ranges from 
roughly $0.8$ to $1.2$.

The calibration marginalization is identical to the general case 
discussed above if we take 
$b=c - 1$ and $\xb = \xp$. 
For clarity in the following 
subsections we will write the calibration uncertainty as $\sigc$, 
but it can be identified with $\sigb$, since $b$ and $c$ have identical 
distributions (though they are about a different mean.)  
Therefore 
\begin{equation}
\M=\N+\sigc^2\xp{\xp}^{\rm T}. 
\label{M_calib} 
\end{equation}
Note that since the the power spectrum measurements, and therefore $c$, 
are positive,
we must have $\sigc \ll 1$ for the Gaussian prior 
for the calibration to be a reasonable approximation. 

Note that this is identical to the formulation used in Wang et al.
(2001) except that they use the approximation that the theory $\xp$ 
is nearly equal to the data $\xo$ to make a theory independent 
correction to the noise matrix.  This is performed just once rather 
than for each theory, making 
it as fast as if the calibration uncertainty was ignored. 
Formally, this is almost equivalent to applying the calibration correction
to the data, but not their error bars.  
In practical terms we find that using $\xo$ instead of $\xp$ 
makes a shift of a few per cent in estimated parameter values.
(The same remarks apply to the beam marginalisation in
the following subsections.)

If we instead attempt to use the exact expression of Eq.~\ref{M_calib} then, 
because the matrix $\M$ (or more importantly, its inverse)  
is a function of the predicted quantities, $\xp$, it must be re-calculated 
for each underlying model and this can be very time consuming.
(Not necessarily as time consuming as 
evaluating the predictions for a given theory, but here we assume 
these are already known.)    
For the simple calibration uncertainty case, a fast computational 
implementation is to calculate in advance the quantities
\begin{eqnarray}
{\bf v}_{\rm o}&=&\N^{-1}\xo\\ \label{fastc_vo}
s_{\rm oo}&=&{\xo}^{\rm T}\N^{-1}\xo \label{fastc_oo}.
\end{eqnarray}
Then,  dropping terms which are independent of the model predictions, 
the effective chi-squared defined by
\begin{eqnarray}
\P(\xo | \N,\, \xp)
&\propto& \exp \left[ -\chi^2_c /2 \right]
\end{eqnarray}
is given by
\begin{eqnarray}
\chi^2_c&=& s_{\rm oo}-2s_{\rm po}+s_{\rm pp}-
(s_{\rm po}-s_{\rm pp})^2 /s_c + \log(s_c \sigc^2)
\label{fastc_chic} \\
s_{\rm po}&=&{\xp}^{\rm T} {\bf v}_1 \label{fastc_spo} \\
s_{\rm pp}&=&{\xp}^{\rm T} \N^{-1} \xp \label{fastc_spp} \\
s_c&=& s_{\rm pp}+\frac{1}{\sigc^2} \label{fastc_sc}.
\end{eqnarray}

If the covariance matrix is not diagonal, the matrix product,
Eq.~\ref{fastc_spp}, scales as 
$n^2$ and this dominates the time required to calculate $\chi^2$. 
If the covariance matrix is diagonal, the matrix product and the vector products
are both linear in $n$ and take comparable amounts of time. 
In either case however, the effects of marginalizing over the 
calibration (the last two terms in Eq.~\ref{fastc_chic}) use the 
same factors required to calculate the naive $\chi^2$, so this 
process takes no more computation time.

\subsection{Beam Uncertainty}

Measurements of the cosmic microwave background radiation angular
power spectrum are often hampered by an uncertainty in the telescope
beam.
This leads to correlated errors in the estimates of power spectrum analogous
to those of the calibration uncertainty, with the added complication that 
the corrections grow as the angular scale becomes smaller. 
Marginalizing over these can be performed in the same way, but now 
$\xb = \A \xp$, where $\A$ is a diagonal matrix which increases for higher 
multipole band powers.  
The precise nature of this matrix depends on the nature of the beam shape and 
`jitter' uncertainties, and is best estimated by the experimental teams. 
For example, 
the BOOMERANG and MAXIMA teams give $1-\sigma$ beam uncertainties, $\dxo$
(for BOOMERANG see Fig.~2 of Netterfield et al. 2001;
for MAXIMA-1 we add in quadrature the beam and pointing
contributions given in Table 1 of Lee et al. 2001).
The matrix $\A$ is given by the percentage errors for each bandpower, that is,
\begin{equation}
\A = {\rm diag}(\d x^{\rm o}_{i}/x^{\rm o}_i)
\label{Abeam}
\end{equation}
where $x^{\rm o}_i$ are the derived bandpowers and the normalization is
such that $\sigma_b = 1$.  
    
The nature of the beam uncertainties, at least on scales larger 
than the beam, can be understood qualitatively by considering the 
case of a Gaussian beam 
whose angular size is not well known.  
For a beam of true size $\theta$ and which has been misestimated to be 
$\theta_0$, the predicted
bandpower $x^{\rm p}_i$ at mean multipole $\ell_i$ must be is
transformed by the relation
\begin{eqnarray}
{x^{\rm p}_i}^{\prime}&=&
x^{\rm p}_i \exp \left[\ell_i^2 (\theta_0^2-\theta^2)\right]\\
&\simeq&
x^{\rm p}_i(1+\ell_i^2(\theta_0^2-\theta^2))
\end{eqnarray}
(assuming the estimate is not too far from the truth.)
Thus if the beam is assumed to be too small, the inferred band power will
be
smaller than the true band power.  The error in the measurement will
increase
for band powers at higher $\ell$.
If the beam uncertainty is small compared to its mean size 
($\sigma_{\theta} << \theta_0 $) and is
Gaussianly
distributed,
then
$(\theta^2 - \theta_0^2)$ will also be Gaussian distributed, with a width
of
$\sigma_{\theta^2} = 2 \sigma_\theta \theta_0$.
The marginalization then can be performed exactly as derived in
Section~\ref{marg}, with
$\sigma_b = \sigma_{\theta^2}$ and $\A = {\rm diag}(\ell_i^2)$, where
$\ell_i$ is the mean multipole of the $i^{\rm th}$ band.
In practice, this is very close the scaling of beam uncertainties given by
the
experiments such as BOOMERANG and MAXIMA.

Fast marginalization over the beam uncertainty alone may be carried
out in a similar way to that over the calibration uncertainty. However,
since the recent experiments with a beam uncertainty also carry a
calibration uncertainty, we proceed straight to marginalization over both
the calibration and beam uncertainties simultaneously.

\subsection{Calibration and Beam Uncertainty}

The equations for fast computation use the quantities already defined
in Eq.s~\ref{fastc_vo}, \ref{fastc_chic} to \ref{fastc_sc} and
\ref{Abeam}.
The calculation can be speeded up by advance calculation of
\begin{eqnarray}
{\bf v}_{a} &=& \A^{\rm T} \N^{-1} \xo\\
{\mathcal Q}_{a}&=&\A^{\rm T} \N^{-1}\\
{\mathcal Q}_{a\, a}&=&\A^{\rm T} \N^{-1} \A.
\end{eqnarray}
The effective chi-squared defined by
\begin{eqnarray}
\P(\xo | \N,\, \xp, \, \sigb, \, \A,\,\sigc)
&\propto& \exp \left[ -\chi^2_{c\, b} /2 \right]
\end{eqnarray}
is given by
\begin{eqnarray}
\chi_{c\,b}^2 &=& 
\chi_{c}^2
-
\frac{(s_{a {\rm o}}-s_{a {\rm p}}-
(s_{\rm po}-s_{\rm pp})s_{a {\rm p}}/s_c)^2}
{1 + s_{a a}-s^2_{a {\rm p}}/s_c} +\log (1+s_{aa}-s_{a {\rm p}}^2/s_c)\\
s_{a {\rm o}}&=&{\xp} ^{\rm T} {\bf v}_{a}\\
s_{a {\rm p}}&=&{\xp} ^{\rm T}{\mathcal Q}_{a} \xp\\
s_{a a}&=&{\xp} ^{\rm T}{\mathcal Q}_{a a} \xp.
\end{eqnarray}
This takes two to three times as long to calculate as when
there are no calibration or beam uncertainties, depending on whether
$\N$ is diagonal (and assuming $n$ is large).

\section{Application to Data}
\label{app}

It has been made clear by the experimental teams that calibration
and (where applicable) beam uncertainties must be included in any
analysis of the data presented. However it is very tempting to 
save considerable computer time by ignoring these uncertainties.
In this section
we apply the formulae presented above to CMB data,
investigating how big a difference the inclusion of calibration
and beam uncertainties make to parameter estimation.
We consider the latest BOOMERANG and COBE data alone,
for simplicity and because this data set has the largest calibration
and beam uncertainties. 
The power spectrum from this experiment was estimated in
$19$ bins spanning the range $75 \le \ell \le 1050$.
Since no information on the BOOMERANG window functions
and full covariance matrix is publicly available yet, we assign a 
top-hat window function for the spectrum in each bin 
and neglect correlations
between bins. This approach is a good approximation 
of the correct one (see de Bernardis et al. 2001) and
does not affect our conclusions.
For our marginalisations over the beam uncertainties,
we took the $1-\sigma$ error bars from Fig.~2 of Netterfield
et al. (2001).
We also include the COBE data using the RADPack packages 
(Dodelson \& Knox 2000). 

The theoretical models are computed using the publicly available 
CMBFAST code (Seljak \& Zaldarriaga). The ranges of our database 
of models, or equivalently top hat priors, are
$0.1<\Omega_{m}<1.0$, $0.015<\Omega_{b}<0.2$,
$0.0<\Omega_{\Lambda}<1.0$ and $0.25<h<0.95$.
We define $h=H_0/(100$~\mbox{$\rm{km} \rm{s}^{-1} \rm{Mpc}^{-1}$}$)$
throughout.
We vary the spectral index of the primordial density perturbations 
within the range $0.5<n_s<1.5$ and we re-scale the amplitude
of fluctuations by a pre-factor $C_{10}$.
We also assume an external Gaussian prior on the
Hubble parameter $h=0.70 \pm 0.1$ and limit the analysis
to models with age $t_0> 10$ Gyrs (see, e.g. Ferreras, Melchiorri,
Silk 2001).
It is important to note that the constraints we will derive on the various
parameters are heavily affected by the size of our database and
by the priors assumed. Considering a background of gravitational
waves or a different optical depth of the universe, for example,
would change our constraints.
Here we illustrate the effect of the CMB systematics on 
just the simplest models.

We find that neglecting the calibration
error the scalar spectral index 
$n_s=0.91 \pm 0.04$, while when including calibration
$n_s=0.89 \pm 0.06$. Even though these numbers are compatible,
it is important to notice that a scale invariant $n_s=1$ 
power spectrum is excluded at $2 \sigma$ in the first case, 
while is still inside two standard deviations when the calibration
error is included.

We consider
the constraints on the physical baryon density parameter $\Omega_bh^2$.
The baryon density plays a crucial role in the determination
of the relative amplitude of the peaks in the power spectrum and
could therefore be significantly affected by beam uncertainty.
This is particularly true if only the first two Doppler peaks are  
well constrained by the data.  (The determination of a third peak would 
likely break this degeneracy.) 
Neglecting beam uncertainty one obtains the tight constraint 
$\Omega_bh^2=0.022 \pm 0.004$, excluding a low 
$\Omega_bh^2 \sim 0.010$ region  which is still compatible with some
Big Bang Nucleosynthesis data at more than $2 \sigma$, while 
including the beam error one infers $\Omega_bh^2=0.020 \pm 0.006$.

We find that the constraint on the universe curvature is insensitive
to whether beam and calibration uncertainties are taken into account,
which is perhaps to be expected since the curvature affects the acoustic
peak positions, rather than their amplitudes.
We also found that the constraints on the cold dark matter
physical density remain at $\Omega_c h^2 = 0.13 \pm 0.05$ 
whether or not calibration and/or beam uncertainties are taken
into account.

The fast method presented in this Paper is crucially important
when combining multiple experiments, each of which may have independent
calibration and beam uncertainties.
These introduce new unknown parameters and current popular methods
are severely affected by an increase in the number of free parameters.
However note that Markov Chain Monte Carlo methods 
(see Knox, Christensen \& Skordis, 2001) are much less affected by 
increases in the number of free parameters. 

Using the formulae presented here, the $\chi^2_{cb}$ values for 
each experiment are simply added together.
Thus the computational time is still only a few times longer than that
when calibration and beam uncertainties are ignored completely.
To illustrate this we combined the latest BOOMERANG, MAXIMA, DASI and
COBE data to obtain cosmological parameter constraints.
We find that the results on the scalar spectral index are $n_s=0.91 \pm 0.02$ 
if we do no take in to account calibration and beam uncertainties, 
while we obtain a similar value but larger error bars $n_s=0.91 \pm 0.04$
when the above systematics are considered.
However, we found $\Omega_{\rm b} h^2 = 0.019^{+0.003}_{-0.002}$
independent of whether calibration and beam uncertainties are
taken into account. This is mainly due to the absence of
beam uncertainty for DASI.

For the BOOMERANG plus COBE analysis we 
also compare with a numerical marginalization,
in which the calibration error is `maximised' over and the beam error
is marginalised over in seven integration steps. We find very
consistent results, thus the analyses carried out by the
experimental teams using this type of numerical approach can 
be relied on. However note that when combining more 
data sets each with independent calibration (and beam) uncertainties
this numerical method is no longer practical.

\section{Conclusion}
\label{discussion}

Our result for the analytic marginalization over calibration uncertainty
is simple, easy to implement and fast. 
In general, a numerical marginalization (integration) 
over calibration and beam
increases the computation time by a factor equal to the number of 
integration steps squared.
Inclusion of calibration uncertainties by analytical methods 
does not increase computation times,  
and adding in beam uncertainties leads 
to a further increase by a factor of only two or three.
This is true irrespective of the number of different data sets, each with
their own independent calibration and beam uncertainties.

We have shown that marginalization over the calibration and beam 
uncertainties can make a significant difference to parameter estimation, 
particularly in widening the error bars on some parameters as much as 
fifty per cent.
We verify that the constraint on the universe curvature is unaffected
by the inclusion of the calibration and/or beam uncertainty,
but that the physical density of baryons or the spectral index of
primordial fluctuations $n_s$ can be significantly affected, 
allowing significantly lower values for $\Omega_{\rm b} h^2$ and widening
the error bars for $n_s$.

In summary, we show that calibration and beam uncertainties 
should be taken into
account and we present a method which allows this to be done
exactly and with little extra work.

\subsection*{ACKNOWLEDGMENTS}
We thank Ofer Lahav, Keith Grainge, 
Adam Ritz, Daniel Eisenstein, Steve Gull and Lloyd Knox for helpful discussions.
We thank the referee for comments which considerably improved the readabilty
of the paper.
SLB, RC and AM acknowledge support from the PPARC.
RK acknowledges support from an EU Marie Curie Fellowship.

\bibliographystyle{/opt/TeX/tex/bib/mn}

\begin{thebibliography}{}

\bibitem{beanm02}
	Bean, R., \& Melchiorri, A. 2002, PRD, 65, 041302 (astro-ph/9911445)
\bibitem{bondcrit01}
   	Bond, J.~R. \& Crittenden, R.~G. 
in {\it Structure Formation in the Universe,} 2001, 
eds. R.~G. Crittenden and N.~G. Turok, p. 241, Dordrecht: Kluwer 
\bibitem{bondjk98}
   	Bond, J.~R., Jaffe, A., \& Knox, L. 1998, PRD, 57, 2117
\bibitem{bondjk00}
   	Bond, J.~R., Jaffe, A., \& Knox, L. 2000, ApJ, 533, 19
\bibitem{bridlezdlhl01}
	Bridle, S.L., Zehavi, I., Dekel, A., Lahav, O., Hobson, M.P., \&
	Lasenby, A.N. 2001, \mnras, 321, 333
\bibitem{debernardisea01}
	de Bernardis, P., et al. 2001, ApJ, in press.(astro-ph/0105296)
\bibitem{dodelsonk99}
	Dodelson, S. \& Knox, L. 2000, Phys.Rev.Lett., 84, 3523
\bibitem{efstathiou00}
	Efstathiou, G. 2000, \mnras, 310, 842 
\bibitem{efstathioub99}
	Efstathiou, G. \& Bond, R. 1999, \mnras, 304, 75
\bibitem{ferrerasms01}
	Ferreras, I., Melchiorri, A. \& Silk, J. 2001, \mnras
~submitted (astro-ph/0105384)
\bibitem{gangargs97}
	Ganga, K., Ratra, B., Gundersen, J., Sugiyama, N. 1997, \apj,
	484, 7
\bibitem{gs98}
        Gawiser E., \&  Silk J., 1998, Science, 280, 1405
\bibitem{Gull} Gull, S.F., 1989, 
in {\it Maximum Entropy and Bayesian Methods}, Cambridge 1988,  
ed. J. Skilling, p. 53, Dordrecht: Kluwer  
\bibitem{jaffeea01}
	Jaffe. A., et al., 2001, PRL, 86, 3475
\bibitem{halversonea01}
	Halverson et al., 2001, ApJ submitted (astro-ph/0104489)
\bibitem{knoxcs01}
	 Knox, L., Christensen, N., Skordis, C., ApJL submitted
(astro-ph/0109232)
\bibitem{kinneymr01}
	Kinney, W., Melchiorri, A. \& Riotto, A. 2001, PRD, 63, 023505
\bibitem{lahavbhs00}
	Lahav, O., Bridle, S.L, Hobson, M.P., Lasenby, A.N., \& Sodre
	Jr., L. 2000, \mnras, 315, L45
\bibitem{lahavea02}
	Lahav, O., et al. 2002, \mnras, submitted (astro-ph/0112162)
\bibitem{langeea01}
        Lange, A., et al. 2001, PRD 53, 042001
\bibitem{ledourdbb00}
	Le Dour, M., Douspis, M., Bartlett, J.G., \& Blanchard,
	A. 2000, A\&A, 364, 369
\bibitem{leeea01}
	Lee, A.T., et al. 2001, preprint (astro-ph/0104459)
\bibitem{lewiscl00}
	Lewis, A., Challinor, A., \& Lasenby, A. 2000, \apj, 538, 473
\bibitem{lineweaver98}
	Lineweaver, C. H. 1998, ApJ, 505, L69
\bibitem{melchiorriea00}
        Melchiorri, A., et al. 2000, \apj, 536, L63
\bibitem{netterfield01}
	Netterfield, C.B., et al. 2001, \apj ~submitted (astro-ph/0104460)
\bibitem{presstvf92}
 	Press, W.~H., Teukolsky, S.~A., Vettering, W.~T., \& Flannery, B.~P.
	1992, Numerical Recipies, Cambridge Univ. Press
\bibitem{pryke}
	Pryke C. et al, \apj ~submitted, 2001 (astro-ph/0104490)
\bibitem{cmbfast}
        Seljak, U., \& Zaldarriaga, M. 1996, \apj, 469, 437
\bibitem{sivia96}
Sivia, D. S. 1996, {\it  Data Analysis A Bayesian Tutorial},
Oxford University Press 
\bibitem{stomporea01}
	Stompor, R., et al., 2001, \apjl ~submitted (astro-ph/0105062)
\bibitem{wangtz01}
	Wang, X., Tegmark, M., Zaldarriaga, M. 2001, PRD submitted 
(astro-ph/0105091)

\end{thebibliography}

\bsp 
\label{lastpage}
\section*{APPENDIX A}

We substitute Eq. \ref{pxgivenb} and \ref{priorb} into Eq. \ref{intdb}
and collect up terms in $b$, assuming $\N^{-1}$ is symmetric.
We then complete the square and use the standard result for the 
integral over a Gaussian.

\begin{eqnarray}
P(\xo | \N, \,\xp,\,\A,\,\sigb) 
&=& 
\frac{N_{\N}}{\sqrt{2 \pi} \sigb}
\int 
\d b
\exp \left[-\frac{1}{2} 
(\xo -(\xp+b\xb))^{\rm T} \N^{-1} 
(\xo -(\xp+b\xb))\right]
\exp \left[ {-\frac{b^2}{2\sigb^2}} \right] \nonumber
\\
&=& 
\frac{N_{\N}}{\sqrt{2 \pi} \sigb}
\int \d b
\exp \left[-\frac{1}{2}
\left(
(\xo -\xp)^{\rm T} \N^{-1} (\xo -\xp)
-2(\xo -\xp)^{\rm T} \N^{-1} \xb b
+\left({\xb}^{\rm T} \N^{-1} \xb +\sigb^{-2} \right)b^2
\right) \right] \nonumber\\
&=&
\frac{N_{\N}}{\sqrt{2 \pi} \sigb}
\exp \left[ -\frac{1}{2} (\xo -\xp)^{\rm T} \N^{-1} (\xo -\xp)\right]\nonumber 
\\
&&
\times \int \d b
\exp \left[ -\frac{1}{2} \left( \left(
\sqrt{{\xb}^{\rm T} \N^{-1} \xb + \sigb^{-2}} b
+\frac{(\xo -\xp)^{\rm T} \N^{-1} \xb}
{\sqrt{{\xb}^{\rm T} \N^{-1} \xb + \sigb^{-2}}}
\right)^2 
-\left(\frac{\left((\xo -\xp)^{\rm T} \N^{-1} \xb \right)^2}
{{\xb}^{\rm T} \N^{-1} \xb + \sigb^{-2}}\right)
\right)\right] \nonumber 
\\
&=&
N_\M \exp \left[ -\frac{1}{2} \left(
(\xo -\xp)^{\rm T} \N^{-1} (\xo -\xp)
-\frac{\left((\xo -\xp)^{\rm T} \N^{-1} \xb\right)^2}
{{\xb}^{\rm T} \N^{-1} \xb + \sigb^{-2}}
\right)
\right], 
\end{eqnarray}
where,
\begin{eqnarray}
N_\M &\equiv& (2\pi)^{-n/2} |\M|^{-1/2} 
= (2\pi)^{-n/2} |\N + \sigb^2 \xb {\xb}^{\rm T}|^{-1/2} \nonumber \\
& = &  N_\N \left(1 + {\xb}^{\rm T} \N^{-1} \xb \sigb^2 \right)^{-1/2} . 
\end{eqnarray}
This result is equivalent to Eq. \ref{result}.

\clearpage
\section*{APPENDIX B} 

Above, we assumed a particular form of $\xpp = \xp + (b-\bar{b}) \xb$, 
where for simplicity we defined $b$ such that $\bar{b} =0$. 
Here we will try to justify this choice, show 
why it often arises and discuss some simple forms that $\xb$ may take.  

In general, the data could be an arbitrary function of the theoretical 
predictions and the nuisance parameter, i.e. $\xpp = \Fb(\xp).$ 
However, often the raw measurements, $\xo$ and $\N$, are not reported directly.
Instead, experimentalists give results which may be compared 
to the theoretical predictions by assuming the best 
estimate of the nuisance parameter, $b = \bar{b}$.  They solve then for 
the inferred observations given that value to find 
${\xo}' \equiv {\Fbb}^{-1} ({\xo})$,  and solve for its noise correlation
matrix in the same way.  
The likelihoods derived using this new variable are equivalent to those 
derived using the original variables.  While the variables are rescaled, so are 
their inferred noise levels, leaving the likelihoods unchanged.  

The theoretical predictions for these new inferred observables are 
\begin{eqnarray}
{\Fbb}^{-1} \Fb(\xp) & = & {\Fbb}^{-1} \Fbb(\xp) + (b-\bar{b}) 
\left. {\partial {\Fbb}^{-1} \Fb(\xp)} \over {\partial b} \right|_{b = \bar{b}} + \ldots 
\nonumber \\ 
& =  & \xp + (b-\bar{b}) \left. {\partial {\Fbb}^{-1} \Fb(\xp)} \over 
{\partial b} \right|_{b = \bar{b}}   + \ldots
\end{eqnarray}
where the first relation follows from a Taylor expansion. 
We can now see that the form of $\xpp$ used above 
arises naturally when we are considering the ${\xo}'$ variables 
and we can associate $\xb$ with a particular function of $\Fb$. 
If the higher order terms become important for reasonable values of $b$, 
then the analytic methods described here will not be applicable. 
This will depend on how well the nuisance parameter is known.  
In addition, it often happens that $\Fb$ is linear in $b$, 
where these higher order terms are 
identically zero.  

Let us consider some simple examples of transformations $\Fb$ and the 
form of the $\xb$ vectors assoiated with them.  If we take a simple 
displacement, $\Fb(\xp) = \xp + b \xf$, where $\xf$ is a fixed template 
independent of the predictions, then it is easy to show that 
${\Fbb}^{-1} \Fb(\xp) = \xp + (b-\bar{b}) \xf$, so $\xb = \xf$. 
We can also consider a simple multiplicative factor 
$\Fb(\xp) =  b \xp$,  as might arise from simple calibration 
uncertainties. 
Then ${\Fbb}^{-1} \Fb(\xp) = \xp + {(b - \bar{b})} \xp/ {\bar{b}} $. 
We can absorb the $1/{\bar{b}}$ into the definition of $b$, so that 
$\xb = \xp$.  
 
Finally, consider the more complicated case where 
$\Fb(\xp) =  \A(b) \xp$, where 
$\A (b) $ is an $n \times n$ matrix.  
Such is the case for 
beam uncertainties, where what is measured are the band powers of the 
map smoothed by the beam and one is attempting to determine 
the band powers of the unsmoothed map.  
In this case, ${\Fbb}^{-1} \Fb(\xp) = \xp + {(b - \bar{b})} 
 {\A}(\bar{b})^{-1}\A'(\bar{b}) \xp $.  Thus, $\xb$ is some 
$n \times n$ matrix times $\xp$.  
In the beam example, if the beam is Gaussian, 
$\A(b) = {\rm diag}(e^{-b \ell_i^2})$, so 
$\xb = - {\rm diag} (\ell_i^2) \xp$.  
In this example, we have ignored higher order terms in $(b - \bar{b})$, 
which may be important for band powers at high $\ell$, where the uncertainties 
caused by the poorly determined beam are of order the size of the measurements.

\end{document}